\documentclass[twocolumn]{aastex631}

\submitjournal{RNAAS}

\shorttitle{Abundances of the A~30 knots}
\shortauthors{Simpson et al.}
\graphicspath{{./}{figures/}}

\begin{document}

\title{Abundance analysis of the J4 equatorial knot of the born-again planetary nebula Abell~30}

\author[0000-0001-6841-0725]{Jordan Simpson}
\email{jordansp@ing.iac.es}
\affiliation{Department of Physics and Astronomy, University of Sheffield, Sheffield, S3 7RH, UK}
\affiliation{Isaac Newton Group of Telescopes, Apartado de Correos 368, E-38700 Santa Cruz de La Palma, Spain}

\author[0000-0003-3947-5946]{David Jones}
\email{djones@iac.es}
\affiliation{Instituto de Astrof\'isica de Canarias, E-38205 La Laguna, Tenerife, Spain}
\affiliation{Departamento de Astrof\'isica, Universidad de La Laguna, E-38206 La Laguna, Tenerife, Spain}

\author[0000-0002-4000-4394]{Roger Wesson}
\email{rw@nebulousresearch.org}
\affiliation{Department of Physics and Astronomy, University College London, Gower Street, London, WC1E 6BT, UK}

\author[0000-0002-6138-1869]{Jorge Garc\'ia-Rojas}
\email{jogarcia@iac.es}
\affiliation{Instituto de Astrof\'isica de Canarias, E-38205 La Laguna, Tenerife, Spain}
\affiliation{Departamento de Astrof\'isica, Universidad de La Laguna, E-38206 La Laguna, Tenerife, Spain}

\begin{abstract}

Abell 30 belongs to a class of planetary nebulae identified as `born-again', containing dense, hydrogen-poor ejecta with extreme abundance discrepancy factors, likely associated with a central binary system. We present intermediate-dispersion spectroscopy of one such feature -- the J4 equatorial knot. We confirm the apparent physical and chemical segregation of the polar and equatorial knots observed in previous studies, and place an upper limit on the abundance discrepancy factor for O$^{2+}$ of 35, significantly lower than that of the polar knots. These findings further reinforce the theory that the equatorial and polar knots originate from different events.
\end{abstract}


\section{Introduction} \label{sec:intro}

The disparity between abundances derived from recombination lines (RLs) and collisionally excited lines (CELs), the ratio of which is known as the abundance discrepancy factor (ADF), was first observed almost 80 years ago \citep{wyse1942}, but its origin is still unclear. 
ADFs are typically 2-3 in planetary nebulae \citep[PNe;][]{gr13}, although in some cases the ADF can reach 2-3 orders of magnitude \citep{wesson03}. Many explanations for these discrepancies have been proposed, including chemical \citep{torres1990}, density \citep{viegas1995}, and temperature inhomogeneities \citep{torres80}, hydrogen-deficient knots \citep{liu2000}, and orbitally-induced temperature resonances \citep{resonance17}. Recently, a link between these high abundance discrepancy PNe and the binarity of their central stars has become evident \citep{wesson18}.

\begin{figure*}
    \centering
    \includegraphics[width=0.38\textwidth]{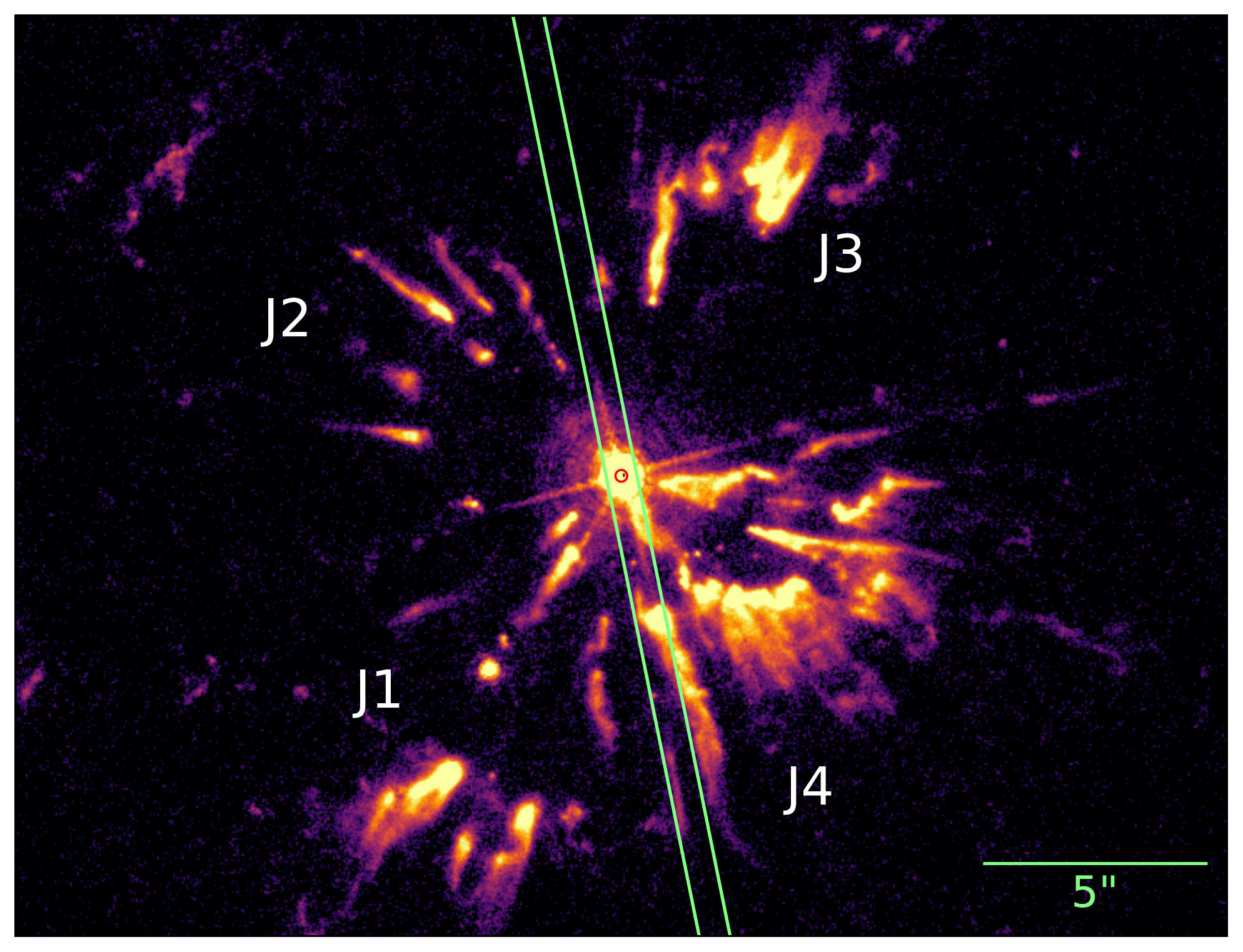}
    \includegraphics[width=0.6\textwidth]{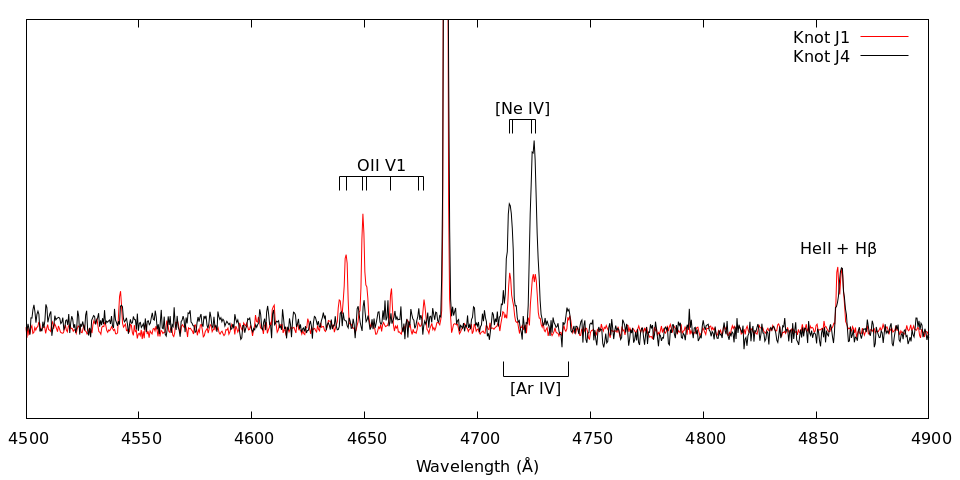}
    \caption{Left: Image of A~30 from \cite{vel14}, with the slit position overlaid. Right: Spectra of the J1 \citep[from][]{wesson03} and J4 (this work) knots scaled to the same H$\beta$ flux.}
    \label{fig:fig}
\end{figure*}

Abell~30 (A~30) presents the highest measured ADF ($\sim$700) in one of its knot complexes \citep{wesson03} and is suspected to host a close binary central star \citep{jacoby20}. A~30 also belongs to a small subset of PNe identified as `born-again', characterised by the presence of low mass, hydrogen-deficient knots. The low C/O ratio measured in the ionised component of the knots led \citet{wesson03} to to argue that they could not have originated in a born-again event, and to suggest a relation to classical novae instead; \citet{lau11} identified scenarios in which the knots could be formed by an eruptive event on a ONeMg white dwarf.  However, \citet{toala21} have recently shown that most of the nebular carbon is in the form of dust and that the total C/O ratio is $>$1 -- more consistent with a very-late thermal pulse scenario.  The knots themselves comprise two components -- a polar jet (J1 and J3), and an equatorial disk (J2 and J4).

Here, we build upon the study of the polar knots by \citet{wesson03} through the analysis of spectra obtained of the equatorial J4 knot of A~30.

\section{Observations and Data Reduction} \label{sec:dr}

A~30 was observed with the Intermediate-dispersion Spectrograph and Imaging System (ISIS) on the 4.2-m William Herschel Telescope (WHT) on February 16 2017.  The setup was replicated from the study of \cite{wesson03}, as were the individual exposure times of 1800s. Six exposures were taken in each arm for a total time on target of 3 hours. The slit was centred on the central star at a position angle of 12$^\circ$, in order to obtain a spectrum of the J4 knot (see Figure \ref{fig:fig}).

\section{Chemical and physical properties} \label{sec:prop}

The extinction, $c(\mathrm{H}\beta)=1.26$, was obtained from the Balmer decrement. This is significantly higher than the value for J1 and J3 \citep{wesson03}, and also differs from the analysis of \cite{guerrero96}, who do not correct for extinction. Ultimately, the extinction depends on the adopted electron temperature (10$^4$~K), and is uncertain in the knots due to their low hydrogen abundances.

The electron density was measured from a weighted mean of the [\ion{O}{2}] $\lambda 3726/ \lambda 3729 $ and [\ion{S}{2}] $\lambda 6716/ \lambda 6732 $ line ratios, giving $n_e = 650 \pm 130 \ \mathrm{cm}^{-3}$.  Adopting this value, the temperature of the knot is then calculated using the [\ion{O}{3}] $(\lambda 4959 + \lambda 5007 / \lambda 4363 $ ratio to be $T_e = 14675 \pm 730 \ \mathrm{K}$. This is in agreement with the value obtained by \cite{guerrero96} of $T_e = 14 000\pm 1000 \ \mathrm{K}$, however the density value obtained is much higher than their value of $n_e = 250 \ \mathrm{cm}^{-3}$.

The relative oxygen abundance was calculated as $12 + \mathrm{\log{O/H}} = 9.72^{+0.27}_{-0.10}$, which is just about within the bounds of the value from \cite{guerrero96} of  $9.51^{+0.11}_{-0.14}$. The relative abundances of nitrogen and helium are in agreement with those of \cite{guerrero96}, with obtained values of $12 + \mathrm{\log{N/H}} = 9.18^{+0.18}_{-0.21}$ and $12 + \mathrm{\log{He/H}} = 12.62^{+0.18}_{-0.17}$. Additionally, relative abundances of Ne, S, Ar$^{2+}$, and Ar$^{3+}$ were measured as $12 + \mathrm{\log{Ne/H}} = 9.25^{+0.18}_{-0.17}$, $12 + \mathrm{\log{S/H}} = 6.96^{+0.27}_{-0.18}$, $12 + \mathrm{\log{Ar^{2+}/H}} = 6.82^{+0.21}_{-0.17}$, $12 + \mathrm{\log{Ar^{3+}/H}} = 7.18^{+0.22}_{-0.17}$. The abundance of C$^{2+}$ was measured as $12 + \mathrm{\log{C^{2+}/H}} = 10.70$ which, assuming C/O$\sim$C$^{2+}$/O$^{2+}$, implies a C/O ratio of 0.5--0.9 consistent with the value foundw for the J1 and J3 knots \citep{wesson03}.


In spite of the high ADF observed for the polar knots, the recombination lines of J4 are very faint (see Fig.~\ref{fig:fig}). The only species we detect both in recombination lines and collisionally excited lines is doubly-ionised oxygen, where there is a low signal-to-noise detection of the \ion{O}{2} lines at $\lambda 4641$ and $\lambda 4649+50$~\AA{}. This detection leads to an approximate $\mathrm{ADF}(\mathrm{O}^{2+})\sim22$. However, the $\lambda 4641$ line is likely contaminated by \ion{N}{3} due to the high excitation of the nebula, and the $\lambda 4649+50$ line appears to be blended with a \ion{C}{3} line. Ultimately, the ADF of J4 is highly uncertain, however assuming no \ion{N}{3} contamination, we derive an upper limit of $\mathrm{ADF}(\mathrm{O}^{2+})=35$.




\section{Conclusions} \label{sec:end}
New spectroscopic observations of the equatorial J4 knot reaffirm the apparent chemical and physical segregation of the polar and equatorial knots of A~30 \citep{guerrero96}.  A low signal-to-noise detection of \ion{O}{2} ORLs places an upper limit on the $\mathrm{ADF}(\mathrm{O}^{2+})=35$, at least one order of magnitude lower than that observed in the polar knots. However, without knowing the mass fraction between the H-poor and H-rich gas phases, the ADF cannot be used as a probe of their relative abundances \citep{gomez-llanos20}. As such, the J4 knot may have similar total abundances to those of the polar knots, but simply present with a different (lower) ratio of H-poor to H-rich material. Ultimately, detailed three-dimensional, bi-phase photo-ionisation modelling will be required in order to further constrain the properties and origins of the H-deficient knots. 



\vspace{5mm}
\facilities{WHT(ISIS), HST}


\software{
          PyNeb \citep{luridiana15}
          ALFA \citep{wesson16}
          NEAT \citep{wesson12}
          }




\bibliography{A30}{}

\begin{thebibliography}{}
\expandafter\ifx\csname natexlab\endcsname\relax\def\natexlab#1{#1}\fi
\providecommand{\url}[1]{\href{#1}{#1}}
\providecommand{\dodoi}[1]{doi:~\href{http://doi.org/#1}{\nolinkurl{#1}}}
\providecommand{\doeprint}[1]{\href{http://ascl.net/#1}{\nolinkurl{http://ascl.net/#1}}}
\providecommand{\doarXiv}[1]{\href{https://arxiv.org/abs/#1}{\nolinkurl{https://arxiv.org/abs/#1}}}

\bibitem[{{Bautista} \& {Ahmed}(2017)}]{resonance17}
{Bautista}, M.~A., \& {Ahmed}, E.~E. 2017, arXiv e-prints, arXiv:1709.07945.
\newblock \doarXiv{1709.07945}

\bibitem[{{Fang} {et~al.}(2014){Fang}, {Guerrero}, {Marquez-Lugo}, {Toal{\'a}},
  {Arthur}, {Chu}, {Blair}, {Gruendl}, {Hamann}, {Oskinova}, \& {Todt}}]{vel14}
{Fang}, X., {Guerrero}, M.~A., {Marquez-Lugo}, R.~A., {et~al.} 2014, \apj, 797,
  100, \dodoi{10.1088/0004-637X/797/2/100}

\bibitem[{{Garc{\'\i}a-Rojas} {et~al.}(2013){Garc{\'\i}a-Rojas}, {Pe{\~n}a},
  {Morisset}, {Delgado-Inglada}, {Mesa-Delgado}, \& {Ruiz}}]{gr13}
{Garc{\'\i}a-Rojas}, J., {Pe{\~n}a}, M., {Morisset}, C., {et~al.} 2013, \aap,
  558, A122, \dodoi{10.1051/0004-6361/201322354}

\bibitem[{{G{\'o}mez-Llanos} \& {Morisset}(2020)}]{gomez-llanos20}
{G{\'o}mez-Llanos}, V., \& {Morisset}, C. 2020, \mnras, 497, 3363,
  \dodoi{10.1093/mnras/staa2157}

\bibitem[{{Gruenwald} \& {Viegas}(1995)}]{viegas1995}
{Gruenwald}, R., \& {Viegas}, S.~M. 1995, \aap, 303, 535

\bibitem[{{Guerrero} \& {Manchado}(1996)}]{guerrero96}
{Guerrero}, M.~A., \& {Manchado}, A. 1996, \apj, 472, 711,
  \dodoi{10.1086/178101}

\bibitem[{{Jacoby} {et~al.}(2020){Jacoby}, {Hillwig}, \& {Jones}}]{jacoby20}
{Jacoby}, G.~H., {Hillwig}, T.~C., \& {Jones}, D. 2020, \mnras, 498, L114,
  \dodoi{10.1093/mnrasl/slaa138}

\bibitem[{{Lau} {et~al.}(2011){Lau}, {De Marco}, \& {Liu}}]{lau11}
{Lau}, H. H.~B., {De Marco}, O., \& {Liu}, X.~W. 2011, \mnras, 410, 1870,
  \dodoi{10.1111/j.1365-2966.2010.17568.x}

\bibitem[{Liu {et~al.}(2000)Liu, Storey, Barlow, Danziger, Cohen, \&
  Bryce}]{liu2000}
Liu, X.-W., Storey, P.~J., Barlow, M.~J., {et~al.} 2000, Monthly Notices of the
  Royal Astronomical Society, 312, 585,
  \dodoi{10.1046/j.1365-8711.2000.03167.x}

\bibitem[{{Luridiana} {et~al.}(2015){Luridiana}, {Morisset}, \&
  {Shaw}}]{luridiana15}
{Luridiana}, V., {Morisset}, C., \& {Shaw}, R.~A. 2015, \aap, 573, A42,
  \dodoi{10.1051/0004-6361/201323152}

\bibitem[{{Toal{\'a}} {et~al.}(2021){Toal{\'a}}, {Jim{\'e}nez-Hern{\'a}ndez},
  {Rodr{\'\i}guez-Gonz{\'a}lez}, {Estrada-Dorado}, {Guerrero},
  {G{\'o}mez-Gonz{\'a}lez}, {Ramos-Larios}, {Garc{\'\i}a-Hern{\'a}ndez}, \&
  {Todt}}]{toala21}
{Toal{\'a}}, J.~A., {Jim{\'e}nez-Hern{\'a}ndez}, P.,
  {Rodr{\'\i}guez-Gonz{\'a}lez}, J.~B., {et~al.} 2021, \mnras, 503, 1543,
  \dodoi{10.1093/mnras/stab593}

\bibitem[{{Torres-Peimbert} {et~al.}(1980){Torres-Peimbert}, {Peimbert}, \&
  {Daltabuit}}]{torres80}
{Torres-Peimbert}, S., {Peimbert}, M., \& {Daltabuit}, E. 1980, \apj, 238, 133,
  \dodoi{10.1086/157966}

\bibitem[{{Torres-Peimbert} {et~al.}(1990){Torres-Peimbert}, {Peimbert}, \&
  {Pena}}]{torres1990}
{Torres-Peimbert}, S., {Peimbert}, M., \& {Pena}, M. 1990, \aap, 233, 540

\bibitem[{{Wesson}(2016)}]{wesson16}
{Wesson}, R. 2016, \mnras, 456, 3774, \dodoi{10.1093/mnras/stv2946}

\bibitem[{{Wesson} {et~al.}(2018){Wesson}, {Jones}, {Garc{\'\i}a-Rojas},
  {Boffin}, \& {Corradi}}]{wesson18}
{Wesson}, R., {Jones}, D., {Garc{\'\i}a-Rojas}, J., {Boffin}, H.~M.~J., \&
  {Corradi}, R.~L.~M. 2018, \mnras, 480, 4589, \dodoi{10.1093/mnras/sty1871}

\bibitem[{{Wesson} {et~al.}(2003){Wesson}, {Liu}, \& {Barlow}}]{wesson03}
{Wesson}, R., {Liu}, X.~W., \& {Barlow}, M.~J. 2003, \mnras, 340, 253,
  \dodoi{10.1046/j.1365-8711.2003.06289.x}

\bibitem[{{Wesson} {et~al.}(2012){Wesson}, {Stock}, \& {Scicluna}}]{wesson12}
{Wesson}, R., {Stock}, D.~J., \& {Scicluna}, P. 2012, \mnras, 422, 3516,
  \dodoi{10.1111/j.1365-2966.2012.20863.x}

\bibitem[{{Wyse}(1942)}]{wyse1942}
{Wyse}, A.~B. 1942, \apj, 95, 356, \dodoi{10.1086/144409}

\end{thebibliography}
\bibliographystyle{aasjournal}



\end{document}